\documentclass[11pt,a4paper]{article}
\usepackage[T1]{fontenc}
\usepackage[utf8]{inputenc}
\usepackage{lmodern,microtype,amsmath,amssymb,amsthm,mathtools}
\usepackage[margin=25mm]{geometry}
\usepackage{booktabs,array,enumitem}
\usepackage[authoryear,round]{natbib}
\usepackage[colorlinks=true,allcolors=blue]{hyperref}
\usepackage{fancyhdr}
\setlist{nosep,leftmargin=*}
\newtheorem{proposition}{Proposition}
\newtheorem{lemma}{Lemma}[section]
\newcommand{\E}{\mathbb E}
\newcommand{\ind}{\mathbf 1}
\newcommand{\doi}[1]{\href{https://doi.org/#1}{\nolinkurl{doi:#1}}}
\hypersetup{pdftitle={Self-Employment as a Career Investment},pdfauthor={Georgy Lukyanov, Konstantin Popov, Shubh Lashkery}}
\title{Self-Employment as a Career Investment}
\author{Georgy Lukyanov\thanks{\raggedright Corresponding author. Toulouse School of Economics, 1 Esp. de l'Universit\'e, 31080 Toulouse, France. Email: \href{mailto:Georgy.Lukyanov@tse-fr.eu}{Georgy.Lukyanov@tse-fr.eu}.}
\and Konstantin Popov\thanks{\raggedright HSE University, International College of Economics and Finance, 11 Pokrovsky Blvd, Moscow, Russian Federation. Email: \href{mailto:kostikrizh@gmail.com}{kostikrizh@gmail.com}.}
\and Shubh Lashkery\thanks{\raggedright CREST--\'Ecole Polytechnique, 5 Av. Le Chatelier, 91120 Palaiseau, France. Email: \href{mailto:shubh.lashkery@polytechnique.edu}{shubh.lashkery@polytechnique.edu}.}}
\date{15 November 2026}
\begin{document}
\maketitle

\begin{abstract}
This paper studies self-employment as an investment in future employment. Workers know their ability and choose between a safe wage and a risky independent project whose outcome becomes public. Employers pay the mean productivity of their applicants. In a two-period model with uniform ability and square-root utility, we establish a range of discount factors on which greater concern for future income increases independent production in both periods and lowers all employment wages. Some workers sacrifice current expected utility to produce a record, then accept employment after either outcome. The record is most valuable to a worker of intermediate ability; the highest types remain independent. We distinguish current income insurance from the cost of giving up this record opportunity. Identifying employers' recruitment cohorts reduces the payment needed to attract some applicant pools, but leaves the equilibrium unchanged on the established range.
\end{abstract}


\section{Introduction}\label{sec:intro}

A worker who takes an independent project receives more than its current income. The project may also provide evidence of what she can do, which prospective employers can use when making an offer. A salaried assignment provides a safer income, but the worker's individual contribution may be difficult to verify outside the firm. Consequently, a worker may choose self-employment even when her ultimate objective is to obtain a job. This paper asks how that choice affects the wages offered to workers who subsequently seek employment.

Consider, for example, an experienced software developer choosing between a salaried assignment within a team and a standalone project whose outcome can be attributed to her. She knows her own ability better than an outside employer does. The independent project exposes her income to risk, but can also provide evidence for a later job application. This comparison applies when project outcomes are verifiable, individual contributions within the team are difficult to establish outside the firm, and the two assignments compete for the worker's time. Employers can still observe where she has worked and infer something from that choice.

We consider a two-period model. In each period, workers can accept a fixed wage or undertake an independent project. Ability is permanent and privately known. The outcome of the first independent project becomes public before the second occupational decision. Firms cannot screen ability directly, and a competitive wage equals the mean productivity of the workers who accept it. The wage therefore reflects both the information in the record and the selection of workers seeking employment with that record.

Our main result concerns the joint response of occupational choices and wages to the weight placed on future income. For the range of discount factors established below, greater concern for the future increases independent production in both periods and lowers every employment wage. Some workers forgo a job that would give them higher current expected utility, produce a record, and return to employment after either success or failure. For these workers, self-employment is a temporary career investment. The highest-ability workers have a different motive: they remain independent regardless of the outcome and obtain no additional value from the later employment option.

The wage response follows from the composition of the workers entering independent production. When future income matters more, the additional producers come from below the original ability cutoff. Their entry lowers the productivity of the pools seeking employment after both outcomes, and hence lowers the wages offered to those pools. The initial employment pool also becomes less productive. Thus the prospect of future employment can induce more independent work even though the resulting employment offers are less generous. Proposition~\ref{thm:regular} establishes these effects jointly, allowing all wages to adjust.

The analysis uses uniform ability and square-root utility. We prove existence, uniqueness within the specified class of wage allocations, and the comparative results for discount factors $0<\beta\leq1/3$. This is a sufficient range for the result. Cutoff behaviour follows more generally, conditional on existence, but the comparative conclusions have not been established for all discount factors. The model is deliberately small enough for the argument to be entirely analytical.

Two further results help interpret the mechanism. Proposition~\ref{prop:record} shows that the prospective record value is greatest at an intermediate ability. We use this result to distinguish the insurance provided by a safe current wage from the compensation required for giving up a future record. Proposition~\ref{prop:credential} allows a new employer's recruits to be publicly identified as a separate cohort. This credential lowers the current payment needed to attract some pools. On the maintained parameter range, however, the reduction is insufficient to make entry profitable, and the equilibrium allocation is preserved.

\subsection{Related literature}

The closest occupational-choice models connect entrepreneurship to information about ability. \citet{HegdeTumlinson2021} show how a discrepancy between private ability and public credentials selects workers into entrepreneurship. \citet{CanidioLegros2023} study task discretion and learning about comparative advantage, including the implications for later careers and the value of failure. \citet{Manso2016} emphasizes experimentation and the option to return to salaried work. Here workers already know their permanent ability. The information produced by an independent project is valuable because it changes employers' assessments, while the wages offered depend on which workers apply. The contribution is the joint response of occupational participation and wages: greater weight on future income expands independent production while lowering employment offers after both outcomes and for workers without a project record. Positive selection into self-employment and the possibility of returning to a job are already familiar mechanisms.

Career concerns link current behaviour to future assessments \citep{Holmstrom1999}. The way information is produced and summarized matters for those incentives. \citet{HornerLambert2021} study the design of ratings to motivate effort, and \citet{BonattiCisternas2020} examine how consumers alter purchases when their histories affect future prices. \citet{CisternasSkills2018} distinguishes incentives to exert effort from incentives to accumulate skills. In our model, ability and the outcome technology are fixed; the worker chooses the occupation that produces a public observation. Reputation also affects experimentation and stopping when the agent learns privately \citep{Thomas2019Experimentation}. Our worker has no such learning problem. Similarly, the private beliefs generated by hidden actions in \citet{CisternasRatchet2018} and \citet{BhaskarMailath2019}, and the learning about match quality in \citet{Jovanovic1979}, are absent here. Our focus is on occupational selection and the wages attached to public outcomes.

Employment can be a source of information in its own right. \citet{Kaniel2025Intermediated} study compensation and career prospects when informed intermediaries certify workers. \citet{StantonThomas2016} document the value of agency affiliation for inexperienced workers in an online labour market. Organizational choices may also affect how credit is assigned for joint output \citep{OnuchicRay2023}. Our cohort comparison isolates the information conveyed by selection into employment, without adding direct screening or employer learning. The fixed wage is itself a restriction on contracts. Unlike \citet{ThomasWorrall1988}, who derive self-enforcing insurance arrangements between a firm and a worker, we take one-period deterministic pay as given and study its interaction with occupational choice.

The empirical literature does not imply a universal hiring advantage from self-employment. \citet{KoellingerMellPohl2015} find fewer employer responses to self-employed applicants; \citet{BotelhoChang2023} find that successful former founders can receive fewer callbacks than failed founders; and \citet{AsanovMavlikeeva2026} document differences across job levels. Subsequent earnings, as studied by \citet{BruceSchuetze2004}, are a separate outcome. These findings caution against reading our wage comparisons as causal effects of having been self-employed. Visible technical work does nevertheless provide a plausible source of career information \citep{LernerTirole2002,AbouElKombozGoldbeck2025}. Open-source activity is an example of such evidence, although it can coexist with a salaried job. The occupational exclusivity in our model is more appropriate for competing assignments that cannot both be undertaken in the same period.

The remainder of the paper is organized as follows. Section~\ref{sec:model} presents the model. Section~\ref{sec:regular} characterizes occupational choices and competitive wages. Section~\ref{sec:insurance} examines the value of a record and its relation to income insurance. Section~\ref{sec:credential} allows employment cohorts to be identified, and Section~\ref{sec:conclusion} concludes. All proofs are in the appendices.

\section{The model}\label{sec:model}

\subsection{Workers and information}

There is a unit mass of workers and two periods, $t=0,1$. Each worker knows her permanent ability $\theta$, drawn uniformly from $[0,1]$. In each period she chooses between employment and independent production. An independent project pays one with probability $\theta$ and zero otherwise; outcomes are independent across periods conditional on ability. Employment pays a deterministic wage $w$ and produces expected output $\theta$ for the firm. The two occupations therefore have the same expected productivity for a given worker.

Workers have utility $u(c)=\sqrt c$ and discount second-period utility by $\beta\in(0,1)$. Income is consumed when earned. There is no saving or separate insurance market. Expected current utility from independent production is $\theta$, whereas a wage $w$ gives utility $\sqrt w$. Either occupation is available in either period.

An independent project's first-period outcome is verified and public: success $S$ or failure $F$. A salaried worker's individual performance does not produce a portable observation. The period and previous occupation are public. A worker who was initially employed therefore has a known employment history without an individual performance outcome, which we call the \emph{blank history}. This history can still reveal ability through selection into employment.

The worker is experienced enough to know her own productivity. The model consequently abstracts from learning one's entrepreneurial aptitude, acquiring skills, and discovering the quality of a particular match. There is no unemployment: a worker who declines employment can always produce independently. Returning to a job is an occupational choice, rather than the resolution of a search spell. The two assignments are mutually exclusive. If the same public evidence could be produced costlessly alongside salaried work, the trade-off we study would disappear.

\subsection{Wages and competition}\label{sec:institution}

Firms are risk neutral and short lived. Each contract offers a fixed wage for one period. Firms cannot screen private ability or select applicants according to unobserved type. Hiring is representative: identical firms receive the same distribution of applicants. Contracts cannot include bonds, screening lotteries, or promises of future employment. Private incumbent information is not carried into a later contract. These assumptions specify the employment institution; the paper does not derive the fixed wage from a richer contracting problem.

In the baseline, all initially employed workers share the blank history. Section~\ref{sec:credential} allows a new firm's recruitment cohort to be identified and priced separately. At every reached second-period history, an active wage must equal applicants' mean productivity and withstand alternative fixed-wage offers.

We consider allocations with a common active initial wage, or no initial offer, and a common active wage at each reached second-period history. Workers choose optimally. Initial competition is tested by allowing a new firm to offer any nonnegative wage and hire a common fraction $\varepsilon$ of all strictly willing applicants. Workers who are not hired retain their fallback: the incumbent offer or independent production. Applicants anticipate the second-period wages generated by the actual recruitment pools.\footnote{Workers indifferent between the entrant and their fallback keep the fallback. Ties between identical incumbents are resolved independently of ability. An entrant therefore cannot choose a favourable set of types from a group of indifferent workers. The common sampling fraction describes its scale and cannot depend on type. Unsuccessful applications are not part of the public record.}

An allocation is an equilibrium in this class if all active wages break even and, for all sufficiently small positive $\varepsilon$, no entrant can profit in any compatible competitive continuation. Compatibility requires workers' choices to generate the pools used to price subsequent employment, with terminal offers satisfying the competition condition above. The bound on entrant size may depend on the baseline and on $\beta$, but must be uniform over its wage offers. The incumbent's initial offer remains fixed during entry; its realized profit need not remain zero after recruitment changes. If there is no initial offer, the entrant creates the employment cohort, whose later wage must be determined from that cohort. There is no commitment to an unused wage quote.

The competition test permits small firms to recruit representative shares of willing workers, with all subsequent prices recalculated. It does not permit a large employer to change a substantial part of the economy or design a different form of contract. Appendix~\ref{app:entry} shows why the test rules out every profitable small offer, including offers arbitrarily close to the incumbent wage.

\section{Occupational choices and wages}\label{sec:regular}

\subsection{The cost of producing a record}

Suppose first-period employment attracts workers with $\theta\leq x$. Competition then sets the initial wage at $x/2$. Let
\begin{equation}
a=\sqrt{x/2}\label{eq:a}
\end{equation}
be the utility of this wage. In the equilibrium characterized below, these workers remain employed in the second period at the same blank-history wage.

A worker accepts a second-period wage $w$ if $\theta\leq\sqrt w$. Denote the employment cutoffs after success and failure by $s$ and $f$, so the corresponding wages are $s^2$ and $f^2$. These wages equal the mean ability of applicants at the respective histories. A worker who has the same record but continues independently is excluded from the wage calculation. Appendix~\ref{app:cutoffs} derives the means and shows that, for each initial cutoff, the two terminal cutoffs are uniquely determined and satisfy $s>f$.

The marginal initial worker takes employment in the second period after either outcome. Her first-period indifference can therefore be written as
\begin{equation}
\underbrace{a-x}_{\text{current utility forgone}}
=\beta\underbrace{\bigl[xs+(1-x)f-a\bigr]}_{\text{expected continuation gain}}.
\label{eq:initialmargin}
\end{equation}
The worker gives up some current utility to improve her later employment option. Both sides depend on wages, which themselves depend on the cutoff. Moreover, indifference and zero-profit pricing alone do not establish an equilibrium: competing offers must also be unprofitable.

\begin{proposition}\label{thm:regular}
Under anonymous employment histories, for $0<\beta\leq1/3$, there is a unique equilibrium allocation in the class of Section~\ref{sec:institution}. The initial employment cutoff satisfies $0<x<1/2$, all employment applicant pools have positive mass, and
\[
x<a<f<s,\qquad f>1/2.
\]
Initial and second-period blank wages equal $x/2$; success and failure wages equal $s^2$ and $f^2$. An increase in $\beta$ lowers the initial cutoff and all employment wages, and raises independent production in both periods. The group $(x,a)$ sacrifices current expected utility to produce a record and takes employment after either outcome. Its mass increases with $\beta$.
\end{proposition}

The proof is in Appendices~\ref{app:cutoffs}--\ref{app:entry}. Wage competition first implies that success receives the higher offer. This ordering yields initial cutoff behaviour for every $\beta<1$, conditional on existence. The bound $1/3$ is used to establish existence, uniqueness, and the comparative results. These conclusions remain restricted to that range; the cutoff argument alone would not establish them.

\subsection{Temporary self-employment}

The equilibrium gives four career paths. Remaining employed is a choice: initial employees can switch to independent production, but prefer the terminal blank-history wage. Indifferent endpoints have zero mass.
\begin{center}
\begin{tabular}{llll}
\toprule
Ability & Initial choice & After success & After failure\\
\midrule
$0<\theta<x$ & Employment & \multicolumn{2}{l}{Blank history: employment}\\
$x<\theta<f$ & Independent & Employment & Employment\\
$f<\theta<s$ & Independent & Employment & Independent\\
$s<\theta<1$ & Independent & Independent & Independent\\
\bottomrule
\end{tabular}
\end{center}
Workers in $(x,f)$ use self-employment temporarily. Those in $(x,a)$ would prefer a job if only current utility mattered, but choose the project for its future return. For workers in $(f,s)$, success improves the employment offer enough to induce a move into a job; failure leaves them working independently. Workers above $s$ remain independent in either case.\footnote{Employment after success and continued independent work after failure can also occur in the myopic limit. The distinctive comparative result is the joint movement of participation and wages when future payoffs receive more weight.}

Giving more weight to the record's continuation gain draws workers from below the original cutoff into independent production. Their entry lowers the mean ability of both outcome pools and the equilibrium wages offered to applicants. The initial employment wage falls as its own pool shrinks. Proposition~\ref{thm:regular} accounts for these wage changes in the occupational decision.

The masses producing independently are
\begin{equation}
P_0=1-x,\qquad P_1=1-f+\frac{f^2-s^2}{2}.\label{eq:P1}
\end{equation}
Both increase as $x$, $s$, and $f$ fall. Since $f>1/2$, every additional initial producer relative to the myopic cutoff $1/2$ subsequently takes a job after either outcome. The smaller group $(x,a)$ identifies workers who also give up current expected utility at the new equilibrium wages.

Although the wage after failure exceeds the blank-history wage, failure is still bad news. Within the original producer pool, it lowers mean ability from $(1+x)/2$ to $(1+2x)/3$. The comparison with blank-history applicants changes the population as well as the information. It cannot be interpreted as a causal hiring reward for failure. This distinction is relevant when comparing the result with evidence on the labour-market outcomes of former founders.

\section{Insurance and the value of a record}\label{sec:insurance}

\subsection{Who values the employment option?}

We can isolate the record's value by holding equilibrium wages fixed. Expected second-period utility after an independent project is
\begin{equation}
B_x(\theta)=\theta\max\{s,\theta\}+(1-\theta)\max\{f,\theta\}.\label{eq:B}
\end{equation}
An initial job instead leaves the worker with the second-period option $\max\{a,\theta\}$. Define the prospective continuation gain from producing a record as
\begin{equation}
\Delta_x(\theta)=B_x(\theta)-\max\{a,\theta\}.\label{eq:Delta}
\end{equation}
This comparison changes one worker's initial occupation, leaving continuation prices unchanged. It measures her incentive, rather than the aggregate effect of changing the information system.

\begin{proposition}\label{prop:record}
In the equilibrium of Proposition~\ref{thm:regular}, $\Delta_x(\theta)$ is positive for $\theta<s$, strictly increasing on $[0,a]$, and strictly decreasing on $[a,s]$. It is zero for $\theta\geq s$. The unique maximum is attained at $\theta=a$.
\end{proposition}

For a worker below $a$, the blank-history employment option is fixed, while greater ability raises the probability of obtaining the better wage after success. Above $a$, the alternative to producing a record also improves with ability: the worker can take an initial job and then work independently. At the top, neither outcome wage induces employment, so the record adds nothing to the worker's future choice. The largest gain therefore accrues to an intermediate type. Appendix~\ref{app:interpretation} gives the piecewise expression and its derivatives under the maintained primitives.

\subsection{Insurance and the reservation wage}

A wage below a worker's expected output need not be an estimate of what she pays for insurance. Independent income has expectation $\theta$ and cash certainty equivalent $\theta^2$. The current cash risk premium is
\begin{equation}
\pi(\theta)=\theta-\theta^2.\label{eq:riskpremium}
\end{equation}
At the final date, a worker accepts employment whenever $w\geq\theta^2$. If $w<\theta\leq\sqrt w$, she accepts less than her expected output in return for certainty. If $\theta<w$, she receives a transfer from more productive applicants. Competition makes these cash transfers sum to zero within the pool; it does not equate each worker's wage to her individual productivity.

Initially, the employment offer must also compensate for the lost record opportunity. Holding continuation prices fixed for an individual worker, the final and initial reservation wages are
\begin{equation}
\rho_1(\theta)=\theta^2,\qquad
\rho_0(\theta)=[\theta+\beta\Delta_x(\theta)]^2.\label{eq:reservation}
\end{equation}
Consequently, the largest acceptable concession below the worker's expected output is
\begin{equation}
\theta-\rho_0(\theta)
=\pi(\theta)-\bigl(\rho_0(\theta)-\rho_1(\theta)\bigr).\label{eq:concession}
\end{equation}
The second term is nonnegative. Giving up the opportunity to produce a record raises the current payment needed to attract the worker. For some low types, the required wage can exceed their own expected output: access to the selected outcome pools makes independent production attractive despite low current productivity. At the initial margin, equation~\eqref{eq:initialmargin} equates the current utility sacrifice to the discounted continuation gain.

These expressions separate three elements of the choice: removing current income risk, receiving or financing a transfer within the applicant pool, and giving up a future employment opportunity. They are accounting comparisons at given prices. A firm recruiting an entire pool must also allow for changes in subsequent wages, as in the competition test of Section~\ref{sec:institution}. They neither assign causal shares to an observed wage gap nor establish a welfare ranking. Appendix~\ref{app:interpretation} provides the corresponding utility and pool identities. Notice also that, for a fixed type who remains employed as $\beta$ rises, the wage falls while $\pi(\theta)$ stays unchanged. The wage response cannot be attributed to a change in that worker's current risk premium.

\section{Employment as a credential}\label{sec:credential}

So far, all initial employees have shared the same subsequent history. We now suppose that outsiders can identify a new firm's recruitment cohort and price its former recruits separately. They observe the recruitment rule, but receive no direct observation of ability or individual salaried performance. Identical baseline employers still recruit representative pools. This modification allows employment to reveal the selection induced by an offer.

Consider an entrant attracting workers in $[0,c]$ as its size tends to zero. Appendix~\ref{app:entry} derives this cutoff form from preferences and representative recruitment. Let $Q_A(c)$ and $Q_L(c)$ be the current payments, in utility units, needed to attract this pool under anonymous and identifiable employment histories. The identifiable cohort's final employment utility is $b(c)=\sqrt{c/2}$ for $c\leq1/2$ and $b(c)=1/2$ for $c\geq1/2$. The difference in required payments is
\begin{equation}
Q_A(c)-Q_L(c)
=\beta\bigl[\max\{b(c),c\}-\max\{a,c\}\bigr].\label{eq:credentialvalue}
\end{equation}
The entrant can offer less today if belonging to its cohort improves the marginal recruit's future option.

\begin{proposition}\label{prop:credential}
For $0<\beta\leq1/3$, identifiable hiring cohorts preserve the unique equilibrium allocation and comparative results of Proposition~\ref{thm:regular}. At that allocation, identifying an entrant's cohort strictly reduces the required current utility payment for $x<c<1/2$. The reduction increases up to $c=a$, then decreases to zero at $c=1/2$. It is zero for $c\geq1/2$.
\end{proposition}

An intermediate recruitment pool benefits from the higher later wage attached to its identity. Even with this advantage, however, attracting a wider pool requires a current wage above that pool's mean productivity. Entry remains unprofitable when future wages are recalculated from the actual recruits. Appendix~\ref{app:entry} proves the result for all sufficiently small entrants, and Appendix~\ref{app:interpretation} derives the payment comparison.

For $c\geq1/2$, the marginal recruit can obtain at least as much second-period utility by working independently under either employment history. The improved employment offer therefore has no value at the recruitment margin. The maximum at $c=a$ concerns a pool cutoff and a payment in utility units; the cash-wage reduction need not have its maximum at the same cutoff.

This comparison covers information conveyed by a hiring cohort. Screening, private employer learning, and skill formation could provide other credentials and alter the allocation. To see the importance of the information restriction, suppose instead that ability were publicly observed before both decisions. Competition would pay each type $w(\theta)=\theta$. Since $\sqrt\theta>\theta$ for every interior type, all would prefer employment. A cohort label supplies much less information than observing ability itself, and Proposition~\ref{prop:credential} does not establish a general monotonic effect of greater signal precision.

\section{Conclusion}\label{sec:conclusion}

Self-employment can be a temporary investment in a future job. In the model, giving greater weight to future payoffs induces more workers to produce a public record. Their entry lowers the mean productivity of the applicant pools and reduces employment wages. The record is most valuable at an intermediate ability; the highest types continue independently after either outcome. Allowing employment cohorts to be identified lowers some recruitment costs, but does not overturn the allocation on the parameter range we establish.

The results concern a specific occupational trade-off under competitive spot wages. They depend on privately known ability, income risk, and a difference in the portability of performance information. Since a given worker has the same expected productivity in either occupation, the increase in independent production leaves expected output at $1/2$ per period. A welfare comparison would require a separate analysis of risk, transfers, and the weight on future consumption. The model identifies an incentive to produce evidence and its effect on wages, without treating either more self-employment or a better individual record as an aggregate productivity gain.

\clearpage
\appendix
\section{Cutoffs and terminal wages}\label{app:cutoffs}

We first derive the cutoff representation used in the main text. For a worker who produced independently at date 0, the outcome changes the distribution of ability. Applying for a job selects that distribution once more. If initial producers have types above $x$ and applicants at an outcome have types below $y$, their conditional means are
\begin{align}
\mu_S(x,y)&=\frac{2(x^2+xy+y^2)}{3(x+y)},\label{eq:muS}\\
\mu_F(x,y)&=\frac{3(x+y)-2(x^2+xy+y^2)}{3(2-x-y)}.\label{eq:muF}
\end{align}
These expressions follow by weighting the uniform density by $\theta$ after success and by $1-\theta$ after failure, and integrating over $(x,y)$. A terminal worker accepts a wage $y^2$ precisely when her type does not exceed $y$. Zero profit therefore requires
\begin{equation}
\mu_S(x,s)=s^2,\qquad \mu_F(x,f)=f^2.\label{eq:terminal}
\end{equation}
Write $a(x)=\sqrt{x/2}$ and define
\begin{equation}
R(x)=xs(x)+(1-x)f(x),\qquad
H(x,\beta)=(1+\beta)a(x)-x-\beta R(x).\label{eq:H}
\end{equation}
Here $R(x)$ is the marginal initial producer's expected terminal utility. The equation $H=0$ is her indifference condition \eqref{eq:initialmargin}. The following arguments show why an equilibrium must have this form.

\begin{lemma}\label{lem:cutoffs}
In any allocation satisfying worker optimality, terminal competition, and the common-active-wage conditions, initial full employment is impossible. Employment has positive mass at every reached terminal history, and outcome wages satisfy $w_S>w_F$. An active common initial wage induces a cutoff for every $\beta<1$. Every allocation with positive initial employment has $0<x<1/2$ and satisfies $H(x,\beta)=0$.
\end{lemma}
\begin{proof}
We first rule out empty employment at a reached terminal history. Its type distribution is nonzero and absolutely continuous. Let $\ell<1$ be the essential lower bound of its support. If $\ell>0$, a wage with utility $q$ just above $\ell$ attracts types near $\ell$ while paying $q^2<\ell$, so it earns a profit.

If $\ell=0$, let $G(q)$ denote the unnormalized mass of types below $q$. Suppose no wage could profitably attract workers. For every sufficiently small $q>0$ we would then have
\[
\int_0^q\theta\,dG(\theta)\leq q^2G(q).
\]
The integral is at least $(q/2)[G(q)-G(q/2)]$, implying $G(q/2)\geq(1-2q)G(q)$. Starting from $0<q_0<1/4$ and iterating gives a strictly positive lower bound on $G(q_0/2^n)$, since the product of the factors $1-2q_0/2^j$ has a positive limit. This contradicts the absence of an atom at zero. An inactive quote does not prevent these deviations: since it attracts no workers, the offers can be chosen above it.

Next suppose all workers were initially employed. The initial wage would be $1/2$. At the terminal uniform blank history, zero profit gives the utility cutoff $q=1/2$ from $q^2=q/2$, and hence wage $1/4$. Types above $1/\sqrt2$ would prefer independent production in both periods to initial employment followed by independent production. Full initial employment is therefore impossible.

Let $h$ be the density of initial independent producers, without yet imposing a cutoff. The success and failure densities are proportional to $\theta h$ and $(1-\theta)h$. At any common truncation with positive mass, denote the mean and variance under truncated $h$ by $m$ and $v$. The difference between the two conditional means is
\[
\mu_S^h-\mu_F^h=\frac{v}{m(1-m)}>0.
\]
If the success wage were below the failure wage, a success-history employer could match the latter and profit from the higher mean of its applicants. Equal wages are inconsistent with zero profit. Thus the utility cutoffs satisfy $s>f$.

For any blank-history utility $b$, put $B(\theta)=\theta\max\{s,\theta\}+(1-\theta)\max\{f,\theta\}$. This function is nondecreasing: both occupational options increase with type, and success gives the weakly better option. The current utility payment required to attract type $\theta$ is
\begin{equation}
T_b(\theta)=\theta+\beta B(\theta)-\beta\max\{b,\theta\},
\qquad T_b'\geq1-\beta>0\quad\hbox{a.e.}\label{eq:singlecross}
\end{equation}
A common initial wage therefore attracts a cutoff pool. This comparison allows a worker to accept the initial job and subsequently produce independently.

An initial cutoff $x>1/2$ would give current utility $\sqrt{x/2}<x$ and terminal blank utility $1/2$. Types just below $x$ would prefer independent production in both periods. At $x=1/2$, the marginal current payoffs coincide, but the record strictly improves continuation utility, again inducing independent production. Hence $0<x<1/2$ whenever initial employment is positive. The uniform blank pool is then fully employed at both dates at wage $x/2$. Lemma~\ref{lem:terminal} shows that the outcome cutoffs exceed $x$, so the marginal worker returns to employment after either outcome. Her indifference is $H=0$.
\end{proof}

\begin{lemma}\label{lem:terminal}
For every $0\leq x<1$, each equation in \eqref{eq:terminal} has a unique positive-pool root in $(x,1)$. The roots increase with $x$, satisfy $s>f>\sqrt x$ for $x>0$, and withstand all terminal fixed-wage offers. At $x=0$,
\begin{equation}
s_*:=s(0)=2/3,\qquad f_*:=f(0)=(4-\sqrt7)/3.\label{eq:limits}
\end{equation}
The competitive terminal utility of a uniform employment cohort $[0,c]$ is $b(c)$ in \eqref{eq:b}.
\end{lemma}
\begin{proof}
For $x<y<1$, differentiation of the conditional means gives
\[
0\leq\partial_y\mu_F\leq1/2,\qquad
\partial_y\mu_S\leq2/3,\qquad
\mu_F\geq y/3,\qquad\mu_S\geq2y/3.
\]
Each ratio $\mu_j(x,y)/y^2$ is therefore strictly decreasing. It starts above one at the lower endpoint and ends below one at $y=1$. When $x=0$, the lower limit is infinite; a zero-pool solution is not a wage allocation. Each positive root is unique and simple. Implicit differentiation yields
\begin{equation}
y_j'(x)=\frac{\partial_x\mu_j(x,y_j)}{2y_j-\partial_y\mu_j(x,y_j)}>0
\quad(x>0),\label{eq:rootderivative}
\end{equation}
since increasing the lower truncation raises mean ability. Moreover,
\[
\mu_S(x,y)-\mu_F(x,y)=\frac{(y-x)^2}{3(x+y)(2-x-y)}>0,
\]
which orders the roots as $s>f$. At either root, $y_j^2=\mu_j>x$. Substitution at $x=0$ gives \eqref{eq:limits}.

A higher offer $z^2$, where $z>y_j$, attracts applicants with mean $\mu_j(x,z)<z^2$ and loses money. A lower offer attracts nobody, while wages above one cannot profit. For a uniform cohort $[0,c]$, all members accept wage $c/2$ if $c\leq1/2$. If $c>1/2$, the terminal cutoff is $y=1/2$, obtained from $y^2=y/2$. The same conditional-mean comparison excludes higher offers in both cases, and Lemma~\ref{lem:cutoffs} excludes empty employment.
\end{proof}

At a root of $H=0$, equation~\eqref{eq:singlecross} with $b=a$ gives $T_a(x)=a$ and verifies every worker's initial choice. The resulting applicant distributions are
\begin{align}
\nu_{E_0}(d\theta)=\nu_{E_1,\mathrm{blank}}(d\theta)
&=x^{-1}\ind_{\{0\leq\theta\leq x\}}\,d\theta,\label{eq:belief0}\\
\nu_{E_1,S}(d\theta)&=\frac{2\theta\ind_{\{x<\theta<s\}}}{s^2-x^2}\,d\theta,\label{eq:beliefS}\\
\nu_{E_1,F}(d\theta)&=\frac{(1-\theta)\ind_{\{x<\theta<f\}}}
{(f-x)-(f^2-x^2)/2}\,d\theta.\label{eq:beliefF}
\end{align}
Every denominator is positive. These are the Bayesian distributions conditional on the history and the decision to apply. No separate belief is chosen at either step. All baseline histories are reached; a new hiring history is priced from the cohort that actually enters it.

\section{The equilibrium root and comparative statics}\label{app:regular}

We locate the roots of the initial indifference equation before imposing initial competition. The argument separates a unique upper root from any lower roots. Appendix~\ref{app:entry} shows that only the upper root survives entry.

For $0\leq x\leq1/5$, the terminal cutoffs and marginal continuation utility satisfy
\begin{equation}
f(x)<1/2+2x/3,\quad s(x)<3/4,\quad
R(x)<U(x):=1/2+11x/12-2x^2/3.\label{eq:upperbounds}
\end{equation}
To obtain the first bound, substitute $y=1/2+2x/3$ into $3(2-x-y)(\mu_F-y^2)$. The resulting polynomial is
\[
N_U(x)=20x^3/9-26x^2/9+11x/12-1/8.
\]
It satisfies $N_U''<0$, $N_U'(1/5)>0$, and $N_U(1/5)<0$. Hence it is increasing and negative on the whole interval, placing the failure root below the proposed bound. The success bound follows from monotonicity and $\mu_S(1/4,3/4)<(3/4)^2$. Weighting the two bounds gives the stated expression for $U$.

For $x\in[1/10,1/5]$ and $\beta\leq1/3$, these inequalities give
\[
H(x,\beta)>g(x):=\frac43\sqrt{x/2}-\frac{47x}{36}+\frac{2x^2}{9}-\frac16.
\]
The lower bound is concave, with positive endpoint values:
\[
g(1/10)=\frac4{3\sqrt{20}}-\frac{59}{200}>0,\qquad
g(1/5)=\frac4{3\sqrt{10}}-\frac{377}{900}>0.
\]
Thus there are no roots in this interval. Since $H(0,\beta)=-\beta f_*<0$ and $H(1/2,\beta)<0$, at least one root lies below $1/10$ and at least one above $1/5$.

To establish uniqueness of the upper root, first note that $f<4/5$ for $x\leq1/2$, by evaluating $\mu_F(1/2,4/5)<(4/5)^2$. Along the failure root,
\[
\partial_x\mu_F\geq1/2,\qquad \partial_y\mu_F\geq11/54.
\]
The second inequality uses $1-x\leq1$, $1-f\geq1/5$, and
$\partial_y\mu_F=\frac23[1-(1-x)^2/(2-x-f)^2]$.
Consequently $2f-\partial_y\mu_F<8/5-11/54<3/2$, and \eqref{eq:rootderivative} implies $f'>1/3$. It follows that
\[
R'=s-f+xs'+(1-x)f'>1/6.
\]
For $x\in[1/5,1/2]$, we have $a'\leq\sqrt{5/8}$ and therefore
\begin{equation}
H_x\leq\frac43\sqrt{5/8}-\frac{19}{18}<0.\label{eq:Hx}
\end{equation}
The bound is obtained by maximizing $(1+\beta)\sqrt{5/8}-1-\beta/6$ over $\beta\leq1/3$. This proves uniqueness above $1/5$.

A stronger lower bound will be useful when checking entry. Direct evaluation gives
\[
\mu_S(2/9,3/4)<(3/4)^2,\qquad
\mu_F(2/9,9/14)<(9/14)^2.
\]
By the decreasing-ratio property in Lemma~\ref{lem:terminal},
$R(2/9)<(2/9)(3/4)+(7/9)(9/14)=2/3$. Since $a(2/9)=1/3$, we obtain $H(2/9,1/3)>0$. Also $H_\beta=a-R<0$. The upper root therefore exceeds $2/9$ for every discount factor in the maintained range. Finally, $x>1/5$ and $\mu_F(1/5,1/2)>(1/2)^2$ imply $f>1/2$.

Once entry selects the upper root, its response to the discount factor follows from implicit differentiation:
\begin{equation}
\frac{dx}{d\beta}=\frac{R-a}{H_x}<0,
\qquad R-a=\frac{a-x}{\beta}>0.\label{eq:comparative}
\end{equation}
Since $s$ and $f$ increase with $x$, both decline with $\beta$, as do all employment wages. The first-period independent mass $1-x$ increases. For the second period, \eqref{eq:P1} gives
$dP_1/d\beta=-s\,ds/d\beta-(1-f)\,df/d\beta>0$.
The mass of workers sacrificing current expected utility satisfies
\[
\frac{d(a-x)}{d\beta}=\left(\frac1{4a}-1\right)\frac{dx}{d\beta}>0,
\]
because $a>1/3$. Every member of this group returns to employment after either outcome, since $a<f$. The same return property holds for $(x,1/2)$, the additional initial producers relative to the myopic allocation, because $f>1/2$. This establishes the comparative and career statements of Proposition~\ref{thm:regular}.

\section{Entry and the wages generated by recruitment}\label{app:entry}

This appendix completes Proposition~\ref{thm:regular} and establishes the allocation claim in Proposition~\ref{prop:credential}. We treat anonymous and identifiable hiring histories together. By Lemma~\ref{lem:terminal}, a uniform employment cohort $[0,c]$ has terminal wage utility
\begin{equation}
b(c)=
\begin{cases}
\sqrt{c/2},&0<c\leq1/2,\\
1/2,&1/2\leq c\leq1.
\end{cases}\label{eq:b}
\end{equation}
We first derive the entrant's applicant pool, then show why its required payment exceeds expected output. The final steps address vanishingly small deviations and exclude the other candidate allocations.

\subsection{The entrant's applicant pool}

Fix an upper-root baseline and keep the incumbent's current utility payment at $a_0=a(x)$. A current entrant payment $q\geq1$ cannot profit, since its cash wage is at least one and mean productivity in a positive-mass pool is strictly below one. We therefore consider $q<1$ and a positive applicant pool. All competitive terminal wage utilities lie in $[0,1]$.

Let $r$ denote the fallback cutoff: among workers not selected by the entrant, types below $r$ take the incumbent job. Terminal outcome competition orders $s>f$ even before any initial cutoff is imposed, so \eqref{eq:singlecross} applies to this fallback decision. Its cutoff is interior. At type zero the reservation utility is at most $\beta<a_0$, whereas at type one it equals one and exceeds $a_0$. The marginal condition also gives
\begin{equation}
0<a_0-\beta\leq r\leq a_0+\beta<5/6.\label{eq:rcompact}
\end{equation}
Indeed, both continuation options lie in $[0,1]$. For sufficiently small entry, a positive fraction of these fallback employees remains with incumbents.

Write $b_e$ and $b_i$ for entrant and incumbent terminal employment utilities. With identifiable cohorts, suppose first that $b_e<b_i$. Positive terminal employment in the entrant cohort requires a willing applicant with $\theta\leq b_e$. Her strict preference implies $q+\beta b_e>a_0+\beta b_i$. The entrant then dominates the incumbent for every type: its disadvantage in continuation utility is never larger than $\beta(b_i-b_e)$. Single crossing against independent production makes its applicant set a prefix $[0,c]$ containing $[0,r]$. The remaining incumbent hires are uniform on $[0,r]$, so $b_i=b(r)$ and $b_e=b(c)\geq b(r)$, a contradiction.

Thus $b_e\geq b_i$. The entrant's advantage over the incumbent is weakly decreasing in type, while its advantage over independent production is strictly decreasing. The intersection of the two strict-preference sets is again a prefix $[0,c]$. It cannot end below $r$. If $c<r$, incumbent hires have density $1-\varepsilon$ below $c$ and one on $(c,r]$, so their conditional mean at every truncation is at least that of a uniform $[0,r]$ cohort. No competitive terminal cutoff can then lie below $b(r)$: for such a cutoff the uniform cohort already has mean productivity above the proposed wage. Hence $b_i\geq b(r)\geq b(c)=b_e$. When $c<1/2$, the last inequality is strict. Otherwise equality requires $b_i=b_e=1/2$, but strict preference for the entrant then requires $q>a_0$ and attracts the entire fallback pool. Both cases contradict $c<r$.

With anonymous histories the two jobs have identical continuations. Any offer attracting workers must pay more than the incumbent and therefore recruits a prefix containing the fallback pool. In either institution, $c=r$ would make the two cohorts' continuations identical. Strict preference for the entrant would require $q>a_0$, inducing some types above $r$ to apply as well. Thus every positive applicant pool has $c>r$. Since $q<1$, type one strictly prefers independent production and $c<1$; marginal-recruit indifference is valid. The tie rule excludes recruitment of a selected subset from a flat indifference set.

\subsection{The limiting recruitment payment}

We first compute the entrant's payment when populated markets are at their baseline limits. The next subsection justifies these limits for every actual small-entry continuation. An identifiable entrant carries a uniform cohort $[0,c]$ into the final period, with utility $b(c)$. Marginal-recruit indifference gives
\begin{align}
Q_A(c)&=c+\beta B_x(c)-\beta\max\{a,c\},\label{eq:QA}\\
Q_L(c)&=c+\beta B_x(c)-\beta\max\{b(c),c\}.\label{eq:QL}
\end{align}
For $c>x$, the entrant's mean current productivity is $c/2$, and its break-even utility payment is $\sqrt{c/2}$.

For $x\leq c\leq1/2$, write $d=s-f$. Since $f>1/2$, $B_x(c)=f+dc$, and the identifiable-cohort payment gap is
\begin{align}
G_L(c)&:=Q_L(c)-\sqrt{c/2}
=(1+\beta d)c+\beta f-(1+\beta)\sqrt{c/2},\label{eq:gap}\\
G_L(x)&=0,\qquad
G_L'(c)=1+\beta d-\frac{1+\beta}{4\sqrt{c/2}}
>1-\frac{1+\beta}{4a}>0.\label{eq:gapslope}
\end{align}
The strict bound uses $a>1/3$ and $\beta\leq1/3$. For $c\geq1/2$, independent production is at least as good as employment in the identified cohort, so
$Q_L(c)=c+\beta[B_x(c)-c]\geq c\geq\sqrt{c/2}$. The inequality is strict at $1/2$ because $f>1/2$, and above $1/2$ because $c>\sqrt{c/2}$. Thus $Q_L(c)^2>c/2$ for every $c>x$. Since $b(c)\geq a$, we also have $Q_A(c)\geq Q_L(c)$. Both forms of limiting entry lose money.

\subsection{All continuations approach the baseline}

The actual producer and anonymous blank-history densities are
\begin{align}
h_{\varepsilon,r,c}(\theta)&=(1-\varepsilon)\ind_{\{r<\theta\leq c\}}+
\ind_{\{c<\theta\leq1\}},\label{eq:hentry}\\
m_{\varepsilon,r,c}(\theta)&=\ind_{\{0\leq\theta\leq r\}}+
\varepsilon\ind_{\{r<\theta\leq c\}}.\label{eq:mentry}
\end{align}
With identifiable histories, incumbent and entrant cohorts are uniform on $[0,r]$ and $[0,c]$. Outcome densities are $\theta h$ and $(1-\theta)h$ in either institution.

Take any sequence of compatible continuations with $\varepsilon\downarrow0$. By \eqref{eq:rcompact}, a subsequence has $r\to\bar r$ and $c\to\bar c$, where $0<\bar r<1$. Terminal employment cannot disappear along this sequence. An outcome wage with utility $y$ prices applicants whose types all exceed $r$, so $y^2>r$ and hence $y>\sqrt r$. On the compact interval in \eqref{eq:rcompact}, the distance $\sqrt r-r$ has a positive lower bound. The densities $h$ are at least $1-\varepsilon$ on $(r,1]$, so outcome applicant masses are bounded away from zero. At the anonymous blank history, a cutoff below $r$ would solve $y^2=y/2$ and equal $1/2$; a cutoff at least $r$ includes mass at least $r$. Its applicant mass therefore cannot vanish either.

We may consequently pass to limits in conditional means. Lemma~\ref{lem:terminal} implies that the outcome utilities converge to $s(\bar r)$ and $f(\bar r)$, and the incumbent continuation utility converges to $b(\bar r)$ under either institution. The limiting fallback condition is
\[
a_0=\bar r+\beta\bigl[R(\bar r)-\max\{b(\bar r),\bar r\}\bigr].
\]
If $\bar r\geq1/2$, its right side exceeds $\bar r\geq1/2>a_0$, since both outcome cutoffs exceed $\bar r$. Thus $\bar r<1/2$, and it must solve
\begin{equation}
K(r):=r+\beta[R(r)-a(r)]=a_0.\label{eq:K}
\end{equation}
This is the relevant equation with the incumbent's initial payment fixed; it does not reimpose initial zero profit after entry.

The only root of \eqref{eq:K} is $r=x$. For $r<1/10$, \eqref{eq:upperbounds} gives $K(r)<1/10+U(1/10)/3=59/200<1/3<a_0$. For $r\in[1/10,1/5]$, Appendix~\ref{app:regular} gives $H(r,\beta)>0$, so $K(r)=a(r)-H(r,\beta)<1/\sqrt{10}<a_0$. For $1/5\leq r<1/2$,
\begin{equation}
K'(r)=1+\beta[R'(r)-a'(r)]>1-\tfrac13\sqrt{5/8}>0.\label{eq:Kprime}
\end{equation}
Since $K(x)=a_0$, no other root is possible. Every convergent subsequence therefore has the same limit. This proves convergence to the baseline for all compatible continuations, uniformly over entrant offers.

\subsection{A uniform bound on the effect of entry}

We now compare the effect of actual recruitment with the strict limiting loss. Let $\ell_S(t)=t$ and $\ell_F(t)=1-t$, and write
\[
D_j(r,y)=\int_r^y\ell_j(t)\,dt,\qquad
N_j(r,y)=\int_r^y t\ell_j(t)\,dt.
\]
For $y>r$, put $v=\min\{c,y\}$. The actual conditional outcome mean is
\begin{equation}
M_j^{\varepsilon}(r,c;y)
=\frac{N_j(r,y)-\varepsilon N_j(r,v)}
{D_j(r,y)-\varepsilon D_j(r,v)}.\label{eq:entrymean}
\end{equation}
For $v>r$, write $D=D_j(r,y)$, $D_c=D_j(r,v)$, $\mu=\mu_j(r,y)$, and $\mu_c=\mu_j(r,v)$. Subtraction gives
\begin{equation}
M_j^{\varepsilon}-\mu
=\frac{\varepsilon D_c(\mu-\mu_c)}{D-\varepsilon D_c}\geq0.\label{eq:entrymeandiff}
\end{equation}
The term is zero when $v=r$. Near the baseline terminal roots, $D$ has a uniform positive lower bound, $D_c\leq c-r$, and $0\leq\mu-\mu_c\leq1$. Hence the mean perturbation is bounded by $C\varepsilon(c-r)$.

These moment bounds control terminal prices as well. At a baseline root, the derivative of $y^2-\mu_j(r,y)$ is $2y-\partial_y\mu_j>0$. On a fixed neighbourhood of the roots it is bounded away from zero. The derivative of an actual conditional mean, wherever it exists, is
\[
\partial_y M_j^{\varepsilon}
=\frac{\ell_j(y)h_{\varepsilon,r,c}(y)
[y-M_j^{\varepsilon}]}{D_j(r,y)-\varepsilon D_j(r,v)}.
\]
Its difference from the baseline derivative is uniformly $O(\varepsilon)$. This bound also holds for both one-sided derivatives at $y=c$. Thus the actual pricing gap remains strictly increasing near each root. Away from the roots its baseline sign is separated from zero on compact sets; the same moment bounds preserve that sign. The terminal roots are unique nearby and withstand all higher-wage offers. Lower offers attract nobody.

For the anonymous blank history, put $u=\min\{r,y\}$ and $v=\min\{c,y\}$. Its truncated mass and first moment are
\[
D_A=u+\varepsilon(v-r)_+,\qquad
N_A=\tfrac12\bigl[u^2+\varepsilon(v^2-r^2)_+\bigr].
\]
Near the baseline, the blank utility $a(r)$ is strictly above $r$. Its applicant mass is positive, and the derivative of the unperturbed gap $y^2-r/2$ is $2y>0$. The same argument bounds the change in its competitive utility by $C\varepsilon(c-r)$. For identified incumbent cohorts the utility is exactly $b(r)=a(r)$ near the baseline. Consequently, if $s_{\varepsilon},f_{\varepsilon},b_i^{\varepsilon}$ denote actual continuation utilities,
\begin{equation}
|s_{\varepsilon}-s(r)|+|f_{\varepsilon}-f(r)|
+|b_i^{\varepsilon}-a(r)|\leq C\varepsilon(c-r).\label{eq:pricecontrol}
\end{equation}
Constants throughout this argument may depend on the fixed baseline and $\beta$, but not on the entrant's offer.

Since $r$ is below all three continuation utilities near the baseline, the actual fallback condition and \eqref{eq:pricecontrol} imply
$|K(r)-a_0|\leq C\varepsilon(c-r)$. Using \eqref{eq:Kprime} gives
\begin{equation}
|r-x|\leq C\varepsilon(c-r)\leq C'\varepsilon|c-x|.\label{eq:rcontrol}
\end{equation}
For the second inequality, use $c-r\leq|c-x|+|r-x|$ and absorb the last term for small $\varepsilon$. If $c<x$, then $r<c<x$ and the first inequality would give $x-r\leq C\varepsilon(c-r)<C\varepsilon(x-r)$, a contradiction. Equality $c=x$ similarly forces $r=x$, which admits no strict recruitment. Thus any actual positive-entry offer has $c>x$.

Let $Q_I^{\varepsilon}(c)$ be its actual required payment, with $I=A,L$. Marginal indifference, the fact that a maximum is Lipschitz in its arguments, and \eqref{eq:pricecontrol}--\eqref{eq:rcontrol} give
\begin{equation}
|Q_I^{\varepsilon}(c)-Q_I(c)|\leq C''\varepsilon(c-x).\label{eq:paymentcontrol}
\end{equation}
By \eqref{eq:gap}--\eqref{eq:gapslope} and continuity, there is a $\kappa>0$ such that $G_L(c)\geq\kappa(c-x)$ on $[x,1]$. Near $x$ use the strictly positive derivative; away from $x$ minimize $G_L(c)/(c-x)$ on a compact interval. Since $Q_A\geq Q_L$, both institutions satisfy
\begin{equation}
Q_I^{\varepsilon}(c)-\sqrt{c/2}
\geq(\kappa-C''\varepsilon)(c-x)>0\label{eq:actualentryloss}
\end{equation}
for sufficiently small $\varepsilon$, uniformly over offers. Every actual entrant's cash wage then exceeds its recruits' mean productivity.

The continuations used in this comparison also exist locally. For any $c>x$, terminal roots are continuous in $(r,c,\varepsilon)$ by the strict gap bounds above. The actual fallback equation differs from $K(r)-a_0$ by at most $C\varepsilon(c-r)$. For a sufficiently large constant $L$, its values at $r=x-L\varepsilon(c-x)$ and $r=x+L\varepsilon(c-x)$ have opposite signs. For sufficiently small $\varepsilon$ this interval lies below $c$, and continuity supplies a fallback root. Set the entrant's payment by marginal indifference at $c$. Since $c>r$, the marginal recruit strictly prefers independent production to the incumbent; the entrant matches the former and dominates the latter. The single-crossing and job-comparison arguments above then implement precisely the prefix $[0,c]$. If $c=1$, a payment of one attracts all interior types and loses money. This proves small-entry robustness without relying on an empty set of compatible continuations.

\subsection{Excluding lower pricing roots}

All other admissible positive pricing roots lie below $1/10$. At such a root, $a<1/\sqrt{20}$ and $d=s-f<1/3$, since $s<3/4$ and $f\geq f_*>5/12$. Set $k=1+\beta d$. Then
\[
4ak<\frac{20}{9\sqrt5}<1.
\]
The anonymous payment gap has derivative $k-1/(4a)<0$ at $c=x$. The identifiable-cohort derivative, $k-(1+\beta)/(4a)$, is smaller still. An offer attracting a fixed pool just above $x$ therefore has strictly positive limiting profit.

This profitable offer can be implemented with small positive entry. At the candidate root, $a=(x+\beta R)/(1+\beta)$ and
\[
1+\beta(R'-a')>1-\frac{\beta}{4a}
\geq1-\frac{1+\beta}{4f_*}>0,
\]
because $4f_*>1+\beta$ throughout the maintained range. Choose the nearby $c>x$ below $a$ and both outcome cutoffs. In this neighbourhood the moment equations are smooth, the fallback equation is locally invertible, and the terminal roots are simple. Their positive higher-offer gaps persist under small entry by the argument leading to \eqref{eq:pricecontrol}. Set the entrant payment by actual marginal indifference at $c$. The resulting fallback cutoff remains below $c$, so the strict-preference comparisons implement the recruited prefix. The strict limiting profit survives the induced changes in recruitment and terminal wages. Lower pricing roots cannot be equilibria.

\subsection{Entry without an initial incumbent}

It remains to exclude the allocation in which every worker initially produces independently. Its terminal outcome cutoffs are $s_*=2/3$ and $f_*=(4-\sqrt7)/3$. Put $d_*=s_*-f_*$ and $k_\beta=1+\beta d_*$. We use the bounds
\[
5/12<f_*<1/2,\qquad 0<d_*<1/4.
\]
There is no incumbent offer. The entrant creates a new employment cohort, and its subsequent wage is determined by that cohort in either information institution.

For a recruited pool with $c\leq f_*$, write $z=\sqrt{c/2}$. Its terminal utility is $b(c)=z$, so the current utility payment required by the marginal recruit is $Q_*(c)=c+\beta[f_*+d_*c-z]$. Subtracting the break-even payment gives
\begin{equation}
Q_*(c)-z=2k_\beta z^2-(1+\beta)z+\beta f_*.
\label{eq:emptygap}
\end{equation}
The minimum of this quadratic occurs at
\[
z_\beta=\frac{1+\beta}{4k_\beta},\qquad c_\beta=2z_\beta^2.
\]
For $0<\beta\leq1/3$, we have $z_\beta\leq1/3$ and $c_\beta\leq2/9<f_*$, so the minimizer belongs to the region used to derive \eqref{eq:emptygap}. The discriminant satisfies
\[
D_\beta=(1+\beta)^2-8\beta f_*k_\beta
>(1+\beta)^2-4\beta(1+\beta/4)
=1-2\beta>0.
\]
The minimized gap is therefore $-D_\beta/(8k_\beta)<0$. Since $f_*>z_\beta$ and $c_\beta>0$, the required payment is itself positive. It follows that $Q_*(c_\beta)^2<c_\beta/2$: the entrant pays less than the average productivity of its recruits. Strict single crossing makes its offer preferable for every type below the marginal recruit.

With small positive entry, the remaining producer density is $1-\varepsilon\ind_{\{\theta<c_\beta\}}$. The simple terminal roots vary continuously with $\varepsilon$. Their pricing gaps remain increasing near the roots and positive at higher cutoffs away from the roots, by the same moment and derivative bounds used above. Thus the perturbed quotes are competitive. Set the offer at the actual reservation payment of $c_\beta$. It attracts the same prefix, except for the indifferent endpoint of zero mass, and retains strictly positive profit for sufficiently small $\varepsilon$. Thus the no-offer allocation is not an equilibrium. Initial full employment was already excluded in Appendix~\ref{app:cutoffs}. Only the upper root remains, completing the allocation claims in Propositions~\ref{thm:regular} and \ref{prop:credential}.

\section{Record value and recruitment payments}\label{app:interpretation}

\subsection{Proof of Proposition~\ref{prop:record}}

Let $d=s-f$. Writing out the occupational options in \eqref{eq:B}--\eqref{eq:Delta} gives
\begin{equation}
\Delta_x(\theta)=
\begin{cases}
f-a+d\theta,&0\leq\theta\leq a,\\
f-(1-d)\theta,&a\leq\theta\leq f,\\
\theta(s-\theta),&f\leq\theta\leq s,\\
0,&s\leq\theta\leq1.
\end{cases}\label{eq:Deltapieces}
\end{equation}
The pieces agree at their boundaries. Their derivatives are $d>0$, $d-1<0$, $s-2\theta$, and zero. Because $f>1/2$ and $s<1$, the third derivative is negative throughout $[f,s]$. The function starts above zero since $f>a$, and remains positive up to $s$, where the quadratic reaches zero. Hence its unique maximum is at $a$.

\subsection{Insurance and pooled transfers}

At a given wage, the current utility gain from employment is
\begin{equation}
\sqrt w-\theta=(\sqrt\theta-\theta)+(\sqrt w-\sqrt\theta).
\label{eq:individual-accounting}
\end{equation}
The first term removes income risk at the worker's own expected output. The second changes that certain payment to the pooled wage. The actuarial benchmark is a comparison, not a contract that must be available when ability is private.

For a fixed competitive applicant pool, $w=\E[\theta\mid\text{pool}]$. Averaging \eqref{eq:individual-accounting} yields
\begin{equation}
\sqrt w-w=\E[\sqrt\theta-\theta\mid\text{pool}]
+\sqrt{\E[\theta\mid\text{pool}]}-\E[\sqrt\theta\mid\text{pool}].
\label{eq:pool-accounting}
\end{equation}
The last difference is a nonnegative Jensen gap. Given the common cardinal utility normalization, it records the effect of redistribution across privately known types. It is not insurance against a worker discovering her own ability. Both equations are accounting identities for a fixed pool.

Initially a worker requires $\sqrt w\geq\theta+\beta\Delta_x(\theta)$; finally she requires $\sqrt w\geq\theta$. Squaring these nonnegative reservation utilities gives \eqref{eq:reservation} and
\begin{equation}
\rho_0(\theta)-\rho_1(\theta)
=2\beta\theta\Delta_x(\theta)+\beta^2\Delta_x(\theta)^2\geq0.
\label{eq:recordcost}
\end{equation}
At the initial margin, the corresponding utility and cash differences are
\[
a-x=\beta\Delta_x(x),\qquad
x/2-x^2=\beta\Delta_x(x)(a+x)>0.
\]
These expressions give the cost of the forgone record opportunity at the prevailing continuation prices.

\subsection{The payment comparison in Proposition~\ref{prop:credential}}

Appendix~\ref{app:entry} established that identifiable cohorts preserve the equilibrium. To obtain the payment comparison, subtract \eqref{eq:QL} from \eqref{eq:QA}. Using \eqref{eq:b}, equation~\eqref{eq:credentialvalue} becomes
\begin{equation}
\eta_x(c):=Q_A(c)-Q_L(c)=
\begin{cases}
\beta[\sqrt{c/2}-a],&x\leq c\leq a,\\
\beta[\sqrt{c/2}-c],&a\leq c\leq1/2,\\
0,&1/2\leq c\leq1.
\end{cases}\label{eq:eta}
\end{equation}
The first piece has derivative $\beta/(4\sqrt{c/2})>0$. The second has derivative $\beta[1/(4\sqrt{c/2})-1]<0$, since $c\geq a>1/3>1/8$. The pieces agree at $a$, and the reduction vanishes at both $x$ and $1/2$. Its unique maximum is $\beta[\sqrt{a/2}-a]$, attained at $c=a$.

Both required payments are positive, so the associated cash-wage reduction is
\[
Q_A(c)^2-Q_L(c)^2=\eta_x(c)[Q_A(c)+Q_L(c)]\geq0.
\]
The sum multiplying $\eta_x(c)$ varies with the pool cutoff. The maximum of the utility-payment reduction therefore need not maximize the cash saving.

\section*{Acknowledgements}
We thank Emiliano Catonini and Steven Kivinen for helpful comments on earlier versions of the paper. We are also grateful to seminar participants at ICEF for useful suggestions. All remaining errors are our own.

\section*{Data availability}
No data were used for the research described in this article.

\end{document}